\let\oldvec\vec
\documentclass[graybox]{svmult}
\let\vec\oldvec

\usepackage{type1cm}        
\usepackage{makeidx}         
\usepackage{graphicx}        
\usepackage{multicol}        
\usepackage[bottom]{footmisc}

\usepackage{newtxtext}       %
\usepackage{newtxmath}       

\usepackage{algorithm}
\usepackage{algorithmicx}
\usepackage{amsmath}
\usepackage{lineno}
\usepackage{multirow}
\usepackage{bm}
\usepackage[utf8]{inputenc}
\usepackage{bbm}
\usepackage{amssymb}
\usepackage{mathrsfs}

\usepackage{etoolbox} 
\usepackage[numbers]{natbib}

\def\eqs{\;}
\newcommand{\map}{\hat{\psi}_i}

\newcommand{\dens}{\texttt{p}}

\makeatletter
\def\ps@pprintTitle{%
 \let\@oddhead\@empty
 \let\@evenhead\@empty
 \def\@oddfoot{}%
 \let\@evenfoot\@oddfoot}
\makeatother

\def\psipop{\psi_{\rm pop}}
\def\iid{\mathop{\sim}_{\rm i.i.d.}}
\newcommand{\trans}[1]{#1^\prime}
\def\thml{\hat{\theta}_{\rm ML}}

\makeindex             

\begin{document}

\title*{Efficient Metropolis-Hastings Sampling for Nonlinear Mixed Effects Models}
\author{Belhal Karimi and Marc Lavielle}
\institute{Belhal Karimi \at Inria, Paris, France, \email{belhal.karimi@inria.fr}
\and Marc Lavielle \at Inria, Paris, France, \email{marc.lavielle@inria.fr}}
%
%
\maketitle

\abstract*{The ability to generate samples of the random effects from their conditional distributions is fundamental for inference in mixed effects models. Random walk Metropolis is widely used to conduct such sampling, but such a method can converge slowly for medium dimension problems, or when the joint structure of the distributions to sample is complex.
We propose a Metropolis–Hastings (MH) algorithm based on a multidimensional Gaussian proposal that takes into account the joint conditional distribution of the random effects and does not require any tuning, in contrast with more sophisticated samplers such as the Metropolis Adjusted Langevin Algorithm or the No-U-Turn Sampler that involve costly tuning runs or intensive computation. Indeed, this distribution is automatically obtained thanks to a Laplace approximation of the original model. We show that such approximation is equivalent to linearizing the model in the case of continuous data. 
Numerical experiments based on real data highlight the very good performances of the proposed method for continuous data model.}

\abstract{The ability to generate samples of the random effects from their conditional distributions is fundamental for inference in mixed effects models. Random walk Metropolis is widely used to conduct such sampling, but such a method can converge slowly for medium dimension problems, or when the joint structure of the distributions to sample is complex.
We propose a Metropolis–Hastings (MH) algorithm based on a multidimensional Gaussian proposal that takes into account the joint conditional distribution of the random effects and does not require any tuning, in contrast with more sophisticated samplers such as the Metropolis Adjusted Langevin Algorithm or the No-U-Turn Sampler that involve costly tuning runs or intensive computation. Indeed, this distribution is automatically obtained thanks to a Laplace approximation of the original model. We show that such approximation is equivalent to linearizing the model in the case of continuous data. 
Numerical experiments based on real data highlight the very good performances of the proposed method for continuous data model.}


\section{Introduction}

Mixed effects models are reference models when the inter-individual variability that can exist within the same population is considered (see \cite{laviellebook} and the references therein).
Given a population of individuals, the probability distribution of the series of observations for each individual depends on a vector of individual parameters. 
For complex priors on these individual parameters or models, Monte Carlo methods must be used to approximate the conditional distribution of the individual parameters given the observations.
Most often, direct sampling from this conditional distribution is impossible and it is necessary to have resort to a Markov chain Monte Carlo (MCMC) procedure.

Designing a fast mixing sampler is of utmost importance for several tasks in the complex process of model building. The most common MCMC method for nonlinear mixed effects models is the \emph{random walk Metropolis} algorithm \cite{casella,robertsoptimal, laviellebook}.
Despite its simplicity, it has been successfully used in many classical examples of pharmacometry, when the number of random effects is not too large.
Nevertheless, maintaining an optimal acceptance rate (advocated in \cite{robertsoptimal}) most often implies very small moves and therefore a very large number of iterations in medium and high dimensions since no information of the geometry of the target distribution is used.

To make better use of this geometry and in order to explore the space faster, the Metropolis-adjusted Langevin algorithm (MALA)
 uses evaluations of the gradient of the target density for proposing new states which are accepted or rejected using the Metropolis-Hastings algorithm \cite{robertsmala, stramer}.
The No-U-Turn Sampler (NUTS) is an extension of the Hamiltonian Monte Carlo \citep{neal2011mcmc} that allows an automatic and optimal selection of some of the settings required by the algorithm, \cite{betancourt2017conceptual}. Nevertheless, these methods may be difficult to use in practice, and are computationally involved, in particular when the structural model is a complex ODE based model. 

The algorithm we propose is a Metropolis-Hastings algorithm, but for which the proposal is a good approximation of the the target distribution.
For general data model (i.e. categorical, count or time-to-event data models or continuous data models), the Laplace approximation of the incomplete pdf $\dens(y_i)$ leads to  a Gaussian approximation of the conditional distribution $\dens(\psi_i |y_i)$.

In the special case of continuous data, linearisation of the model leads, by definition, to a Gaussian linear model for which the conditional distribution of the individual parameter $\psi_i$ given the data $y_i$ is a multidimensional normal distribution that can be computed and we fall back on the results of \citep{karimi2017vi}.

\section{Mixed Effect Models}
\label{sec:mixed}
\subsection{Population Approach and Hierarchical Models}

We will adopt a population approach in the sequel, where we consider $N$ individuals and $n_i$ observations for individual $i$. The set of observed data is $y = (y_i, 1\leq i \leq N)$ where $y_i = (y_{ij}, 1\leq j \leq n_i)$ are the observations for individual $i$. For the sake of clarity, we assume that each observation $y_{ij}$ takes its values in some subset of $\mathbb{R}$. The distribution of the $n_i-$vector of observations $y_i$ depends on a vector of individual parameters $\psi_i$ that takes its values in a subset of $\mathbb{R}^{p}$.
We assume that the pairs $(y_i,\psi_i)$ are mutually independent and consider a parametric framework: the joint distribution of $(y_i,\psi_i)$ is denoted by $\dens(y_i,\psi_i;\theta)$, where $\theta$ is the vector of fixed parameters of the model. 
A natural decomposition of this joint distribution writes $\dens(y_i,\psi_i;\theta) = \dens(y_i|\psi_i;\theta)\dens(\psi_i;\theta)$, where $\dens(y_i|\psi_i;\theta)$ is the conditional distribution of the observations, given the individual parameters, and where $\dens(\psi_i;\theta)$ is the so-called population distribution used to describe the distribution of the individual parameters within the population. 
A particular case of this general framework consists in describing each individual parameters $\psi_i$ as a typical value $\psipop$, and a vector of individual random effects $\eta_i$: $\psi_i = \psipop+\eta_i$.
In the sequel, we will assume a multivariate Gaussian distribution for the random effects: $\eta_i \iid \mathcal{N}(0,\Omega)$. 
Several extensions of this model are straightforward, considering for instance transformation of the normal distribution, or adding individual covariates in the model.

\subsection{Continuous Data Models}  \label{sec:cont}

A regression model is used to express the link between continuous observations and individual parameters:
\begin{equation}\label{continuousmodel}
y_{ij} = f(t_{ij},\psi_i) + \varepsilon_{ij}\eqs,
\end{equation}
where $y_{ij}$ is the j-th observation for individual $i$ measured at time $t_{ij}$, $\varepsilon_{ij}$ is the residual error, $f$ is the structural model assumed to be a twice differentiable function of $\psi_i$. 
We start by assuming that the residual errors are independent and normally distributed with zero-mean and a constant variance $\sigma^2$. Let $t_i=(t_{ij}, 1\leq n_i)$ be the vector of observation times for individual $i$. Then, the model for the observations rewrites
$
 y_i|\psi_i \sim \mathcal{N}(f_i(\psi_i),\sigma^2\texttt{Id}_{n_i\times n_i})\eqs,
$
where $f_i(\psi_i) = (f(t_{i,1},\psi_i), \ldots , f(t_{i,n_i},\psi_i))$.
If we assume that $\psi_i \iid \mathcal{N}(\psipop,\Omega) $, then the parameters of the model are $\theta= (\psipop, \Omega, \sigma^2)$.

\section{Sampling from Conditional Distributions}

The conditional distribution $\dens(\psi_i | y_i ; \theta)$ plays a crucial role in most methods used for inference in nonlinear mixed effects models.

One of the main task to perform is to compute the maximum likelihood (ML) estimate of $\theta$, $\thml = \arg \max \limits_{\theta \in \Theta} {\cal L}(\theta , y)$, where ${\cal L}(\theta , y) \triangleq \log \dens(y;\theta)$.
The stochastic approximation version of EM \cite{lavielle} is an iterative procedure for ML estimation that requires to generate one or several realisations of this conditional distribution at each iteration of the algorithm.


Metropolis-Hasting algorithm is a powerful MCMC procedure widely used for sampling from a complex distribution \citep{brooks2011handbook}.
To simplify the notations, we remove the dependency on $\theta$.
For a given individual $i$, the MH algorithm, to sample from the conditional distribution $\dens(\psi_i | y_i)$, is described as:
\begin{algorithm}[H]
\textbf{Initialization}: Initialize the chain sampling $\psi_i^{(0)}$ from some initial distribution $\xi_i$ .\\
\textbf{Iteration k}: given the current state of the chain $\psi_i^{(k-1)}$:
\begin{enumerate}
\item Sample a candidate $\psi_i^c$ from a proposal distribution $q_i( \, \cdot \, | \psi_i^{(k-1)})$.
\item Compute the MH ratio:
\begin{equation}
\alpha(\psi_i^{(k-1)},\psi_i^{c}) = 
\frac{\dens(\psi_i^{c}|y_i)}{\dens(\psi_i^{(k-1)}|y_i)}
\frac{q_i(\psi_i^{(k-1)}|\psi_i^c)}{q_i(\psi_i^{c}|\psi_i^{(k-1)}) }\eqs.
\end{equation}
\item Set $\psi_i^{(k)}=\psi_i^c$ with probability $\min (1,\alpha(\psi_i^{c},\psi_i^{(k-1)})$ (otherwise, keep $\psi_i^{(k)}=\psi_i^{(k-1)}$).
\end{enumerate}
\caption{Metropolis-Hastings algorithm}
\label{alg:mh}
\end{algorithm}

Current implementations of the MCMC algorithm, to which we will compare our new method, in Monolix \cite{chan}, saemix (R package) \cite{comets}, nlmefitsa (Matlab) and NONMEM \cite{beal} mainly use the same combination of proposals. 
The first proposal is an independent Metropolis-Hasting algorithm which consists in sampling the candidate state directly from the marginal distribution of the individual parameter $\psi_i$. The other proposals are component-wise and block-wise  random walk procedures \cite{metropolis} that updates different components of $\psi_i$ using univariate and multivariate Gaussian proposal distributions. 
Nevertheless, those proposals fail to take into account the nonlinear dependence structure of the individual parameters.
A way to alleviate these problems is to use a proposal distribution derived from a discretised Langevin diffusion whose drift term is the gradient of the logarithm of the target density leading to the Metropolis Adjusted Langevin Algorithm (MALA) \citep{robertsmala, stramer}. The MALA proposal is given by:
\begin{equation}\label{eq:update.mala}
 \psi_i^c \sim \mathcal{N}(\psi_i^{(k)}-\gamma \nabla_{\psi_i} \log \dens (\psi_i^{(k)}|y_i),2\gamma)\eqs,
\end{equation}
where $\gamma$ is a positive stepsize. 
These methods still do not take into consideration the multidimensional structure of the individual parameters. 
Recent works include efforts in that direction, such as the Anisotropic MALA for which the covariance matrix of the proposal depends on the gradient of the target measure \citep{kuhnamala}.
The MALA algorithm is a special instance of the Hybrid Monte Carlo (HMC), introduced in \citep{neal2011mcmc}; see \citep{brooks2011handbook} and the references therein, and consists in augmenting the state space with an auxiliary variable $p$, known as the velocity in Hamiltonian dynamics.

All those methods aim at finding the proposal $q$ that accelerates the convergence of the chain. Unfortunately they are computationally involved and can be difficult to implement (stepsizes and numerical derivatives need to be tuned and implemented).

We see in the next section how to define a multivariate Gaussian proposal for both continuous and noncontinuous data models, that is easy to implement and that takes into account the multidimensional structure of the individual parameters in order to accelerate the MCMC procedure.

\section{A Multivariate Gaussian Proposal}
For a given parameter value $\theta$, the MAP estimate, for individual $i$, of $\psi_i$ is the one that maximises the conditional distribution $\dens(\psi_i|y_i,\theta)$: 
$$\map  = \arg \max \limits_{\psi_i} \dens(\psi_i|y_i,\theta) = \arg \max \limits_{\psi_i} \dens(y_i|\psi_i,\theta)\dens(\psi_i,\theta)$$
\subsection{General Data Models}
For both continuous and noncontinuous data models, the goal is to find a simple proposal, a multivariate Gaussian distribution in our case, that approximates the target distribution $\dens(\psi_i|y_i)$.
In our context, we can  write the marginal pdf $\dens(y_i)$ that we aim to approximate as $\dens(y_i) = \int{e^{\log \dens(y_i,\psi_i)}\textrm{d}\psi_i}$. Then, the Taylor expansion of $\log (\dens(y_i,\psi_i)$ around the MAP $\map$ (that verifies by definition
 $\nabla \log \dens(y_i,\map)=0$) yields the Laplace approximation of $-2\log (\dens(y_i))$ as follows:
$$
-2\log \dens(y_i)  \approx -p\log2\pi - 2\log \dens(y_i,\map) + \log \left(\left|-\nabla^2 \log \dens(y_i,\map)\right|\right)\eqs.
$$
We thus obtain the following approximation of $\log \dens(\map|y_i)$:
 $$
\log \dens(\map|y_i) \approx -\frac{p}{2}\log2\pi  -\frac{1}{2}\log \left(\left|-\nabla^2 \log \dens(y_i,\map) \right|\right)\eqs,
$$
which is precisely the log-pdf of a multivariate Gaussian distribution with mean $\map$ and  variance-covariance $-\nabla^2 \log \dens(y_i,\map)^{-1}$, evaluated at $\map$.

\begin{proposition} \label{lemma:noncont}
The Laplace approximation of the conditional distribution $\psi_i|y_i$ is a multivariate Gaussian distribution with mean $\map$ and variance-covariance 
$$ \Gamma_i =-\nabla^2 \log \dens(y_i,\map)^{-1}= \left(- \nabla^2 \log \dens(y_i|\map)+\Omega^{-1}\right)^{-1} \eqs.$$
\end{proposition}


We shall now see another method to derive a Gaussian proposal distribution in the specific case of continuous data models.

\subsection{Nonlinear Continuous Data Models} \label{mh:nonlinear}
When the model is described by \eqref{continuousmodel}, the approximation of the target distribution can be done twofold: either by using the Laplace approximation, as explained above, or by linearizing the structural model $f_i$ for any individual $i$ of the population.
Once the MAP estimate $\map$ has been computed, using an optimisation procedure, the method is based on the linearisation of the structural model $f$ around $\map$:
\begin{equation}
f_i(\psi_i) \approx f_i(\map) + \textrm{J}_{f_i(\map)}(\psi_i - \map)\eqs,
\end{equation}
where $\textrm{J}_{f_i(\map)}$ is the Jacobian matrix of the vector $f_i(\map)$.
Defining $z_i \triangleq y_i - f_i(\map) + \textrm{J}_{f_i(\map)} \map$ yields a linear model $z_i = \textrm{J}_{f_i(\map)}\psi_i + \epsilon_i$ which tractable conditional distribution can be used for approximating $\dens(\psi_i|y_i,\theta)$:

\begin{proposition} \label{lemma:cont}
Under this linear model, the conditional distribution $\psi_i|y_i$ is a Gaussian distribution with mean $\mu_i$ and variance-covariance $\Gamma_i$ where
\begin{align}
\mu_i = \map \quad \textrm{and} \quad \Gamma_i=\left( \frac{ \trans{\textrm{J}_{f_i(\map)}}  \textrm{J}_{f_i(\map)}}{\sigma^2} + \Omega^{-1}\right)^{-1}\eqs.
\end{align}
\end{proposition}
We can note that linearizing the structural model is equivalent to using the Laplace approximation with the expected information matrix. Indeed:
\begin{equation}\label{eq:expectedfim}
 \mathbb{E}_{y_i|\map}\left(- \nabla^2 \log \dens(y_i|\map) \right) =  \frac{\trans{\textrm{J}_{f_i(\map)}} \textrm{J}_{f_i(\map)} }{\sigma^2}\eqs.
\end{equation}
We then use this normal distribution as a proposal in algorithm \ref{alg:mh} for model \eqref{continuousmodel}.
\section{A Pharmacokinetic Example}\label{sec:numericalexamples}
\subsection{Data and Model}

32 healthy volunteers received a 1.5 mg/kg single oral dose of warfarin, an anticoagulant normally used in the prevention of thrombosis \cite{oreilly}, for who  we measure warfarin plasmatic concentration at different times.
We will consider a one-compartment pharmacokinetics (PK) model for oral administration, assuming first-order absorption and linear elimination processes:
\begin{equation} \label{pkmodel}
f(t,ka, V, k) = \frac{D\,ka}{V(ka - k)}(e^{-ka\,t}-e^{-k\,t})\eqs,
\end{equation}
where $ka$ is the absorption rate constant, $V$ the volume of distribution , $k$ the elimination rate constant, and $D$ the dose administered. 
Here, $ka$, $V$ and $k$ are  PK parameters that can change from one individual to another. 
Then, let $ \psi_i=(ka_i, V_i, k_i)$ be the vector of individual PK parameters for individual $i$ lognormally distributed.
We will assume in this example that the residual errors are independent and normally distributed with mean 0 and variance $\sigma^2$.
We can use the proposal given by Proposition~\ref{lemma:cont} and based on a linearisation of the structural model $f$ proposed in \eqref{pkmodel}.
For the method to be easily extended to any structural model, the gradient is calculated by automatic differentiation using the R package `Madness' \citep{pav2016madness}.
\subsection{MCMC Convergence Diagnostic}
\begin{figure}[thp]
\begin{center}
\includegraphics[width=\textwidth]{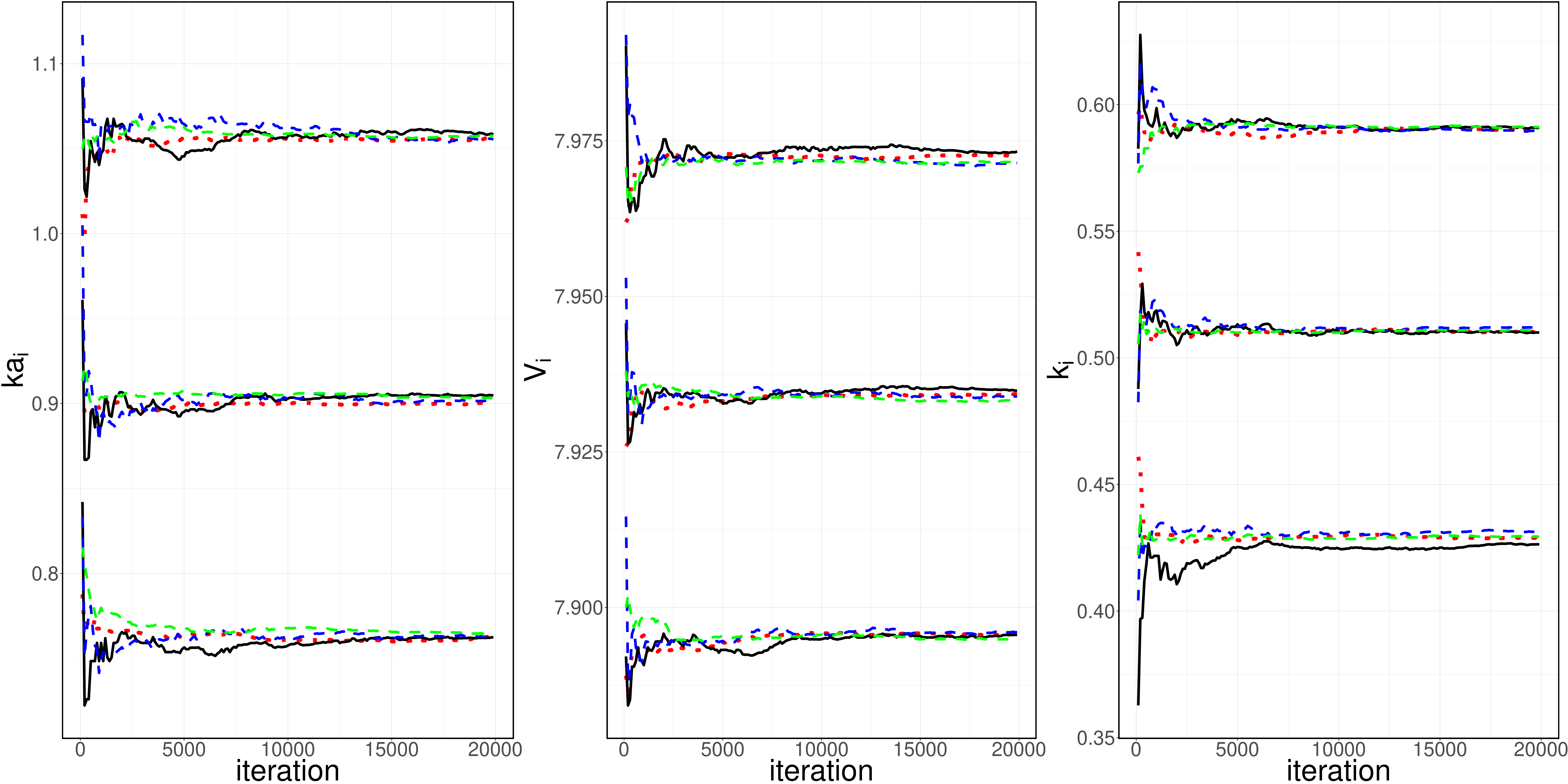}
\caption{Modelling of the warfarin PK data: convergence of the empirical quantiles of order 0.1, 0.5 and 0.9 of $\dens(\psi_i | y_i ; \theta)$ for a single individual. Our new MH algorithm is in red and dotted, the RWM is in black and solid, the MALA is in blue and dashed and the NUTS is in green and dashed.}
\label{q_malanuts}
\end{center}
\end{figure}

We will consider one of the 32 individuals for this study and fix $\theta$ to some arbitrary value, close to the Maximum Likelihood (ML) estimate obtained with SAEM (saemix R package \cite{comets}):
$ka_{\rm pop} =1$, $V_{\rm pop}= 8$, $k_{\rm pop}=0.01$,  $\omega_{ka}=0.5$, $\omega_{V}=0.2$, $\omega_{k}=0.3$ and $\sigma^2=0.5$.
First, we compare our our nlme-IMH, which is a MH sampler using our new proposal, with the RWM, the MALA, which proposal, at iteration $k$, is defined by $\psi_i^c \sim \mathcal{N}(\psi_i^{(k)}-\gamma_k \nabla \log \pi (\psi_i^{(k)}),2\gamma_k)$. The stepsize ($\gamma = 10^{-2}$) is constant and is tuned such that the optimal acceptance rate of $0.57$ is reached \cite{robertsoptimal}. $20\,000$ iterations are run for each algorithm.
Figure~\ref{q_malanuts} highlights quantiles stabilisation using the MALA similar to our method for all orders and dimensions.
The NUTS, implemented in rstan (R Package \cite{rstan}), is fast and steady and presents similar, or even better convergence behaviors for some quantiles and dimension, than our method (see Figure~\ref{q_malanuts}).

\begin{figure}[thp]
\begin{center}
\includegraphics[width=\textwidth]{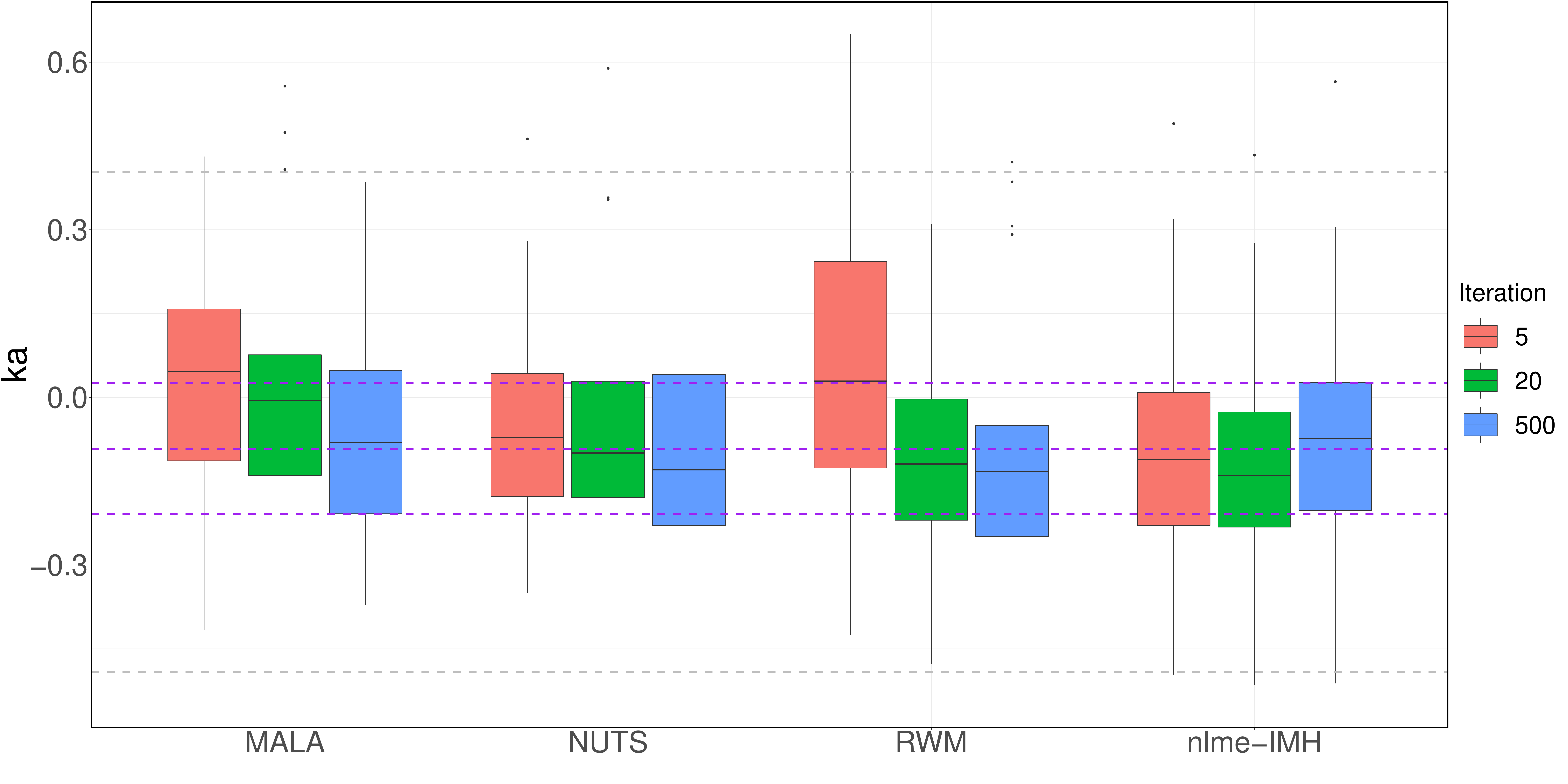}
\caption{Modelling of the warfarin PK data: Boxplots, over $100$ parallel runs, for the RWM, the nlme-IMH, the MALA and the NUTS algorithm. The ground truth median, $0.25$ and $0.75$ percentiles are plotted as a dashed purple line and its maximum and minimum as a dashed grey line.}
\label{cont:boxplots}
\end{center}
\end{figure}

Then, we produce $100$ independent runs of the RWM, the IMH using our proposal distribution (called the nlme-IMH algorithm), the MALA and the NUTS for $500$ iterations. The boxplots of the samples drawn at a given iteration threshold are presented Figure ~\ref{cont:boxplots} against the ground truth (calculated running the NUTS for $100\,000$ iterations) for the parameter \textbf{ka}.

For the three numbers of iteration considered in Figure \ref{cont:boxplots}, the median of the nlme-IMH and NUTS samples are closer to the ground truth. Figure ~\ref{cont:boxplots} also highlights that all those methods succeed in sampling from the whole distribution after $500$ iterations.
Similar comments can be made for the other parameters.

We decided to conduct a comparison between those sampling methods just in terms of number of iterations (one iteration is one transition of the Markov Chain). We acknowledge that the transition cost is not the same for each of those algorithms, though, our nmle-IMH algorithm, except the initialisation step that requires a MAP and a Jacobian computation, has the same iteration cost as RWM. The call to the structural model $f$ being very costly in real applications (when the model is the solution of a complex ODE for instance), the MALA and the NUTS, computing its first order derivatives at each transition, are thus far computationally involved.

Since computational costs per transition are hard to accurately define for each sampling algorithm and since runtime depends on the actual implementation of those methods, comparisons are based on the number of iterations of the chain here.

\section{Conclusion and Discussion}
We presented in this article an independent Metropolis-Hastings procedure for sampling random effects from their conditional distributions in nonlinear mixed effects models. 
The numerical experiments that we have conducted show that the proposed sampler converges to the target distribution as fast as state-of-the-art samplers. 
This good practical behaviour is partly explained by the fact that the conditional mode of the random effects in the linearised model coincides with the conditional mode of the random effects in the original model. 
Initial experiments embedding this fast and easy-to-implement IMH algorithm within the SAEM algorithm \cite{lavielle}, for Maximum Likelihood Estimation, indicate a faster convergence behavior.

\newpage
\bibliographystyle{apa}
\bibliography{ref}
\end{document}